\documentclass[11pt, a4paper]{article}

\usepackage{amsthm}
\usepackage{amsmath}
\usepackage{amsfonts}
\usepackage{booktabs}
\usepackage{natbib}
\usepackage{hyperref}
\usepackage{graphicx}
\usepackage{color}
\usepackage[left = 2.5cm, right = 2.5cm, bottom = 3cm, top = 2.5cm]{geometry}
\usepackage{dcolumn}
\usepackage{authblk}

\newtheoremstyle{example}
{3pt} 
{3pt} 
{} 
{0\parindent} 
{\bf}
{:} 
{.5em} 
{} 
\newtheoremstyle{theorem}
{3pt} 
{3pt} 
{\em} 
{0\parindent} 
{\bf}
{:} 
{.5em} 
{} 
\theoremstyle{example} 
\theoremstyle{theorem} 
\theoremstyle{theorem} \newtheorem{conjecture}{Conjecture}[section]

\def\expect{{\mathop{\rm E}}}
\def\var{{\mathop{\rm var}}}

\title{Jeffreys-prior penalty for high-dimensional logistic regression: A
  conjecture about aggregate bias}

\author[1]{Ioannis Kosmidis\thanks{ioannis.kosmidis@warwick.ac.uk}}
\author[1]{Patrick Zietkiewicz\thanks{patrick.zietkiewicz@warwick.ac.uk}}

\affil[1]{Department of Statistics, University of Warwick \authorcr Gibbet Hill Road, Coventry, CV4 7AL, UK}

\begin{document}

\maketitle

\begin{abstract}
  Firth (1993, Biometrika) shows that the maximum Jeffreys' prior
  penalized likelihood estimator in logistic regression has asymptotic
  bias decreasing with the square of the number of observations when
  the number of parameters is fixed, which is an order faster than the
  typical rate from maximum likelihood. The widespread use of that
  estimator in applied work is supported by the results in Kosmidis
  and Firth (2021, Biometrika), who show that it takes finite values,
  even in cases where the maximum likelihood estimate does not exist.
  Kosmidis and Firth (2021, Biometrika) also provide empirical
  evidence that the estimator has good bias properties in
  high-dimensional settings where the number of parameters grows
  asymptotically linearly but slower than the number of
  observations. We design and carry out a large-scale computer
  experiment covering a wide range of such high-dimensional settings
  and produce strong empirical evidence for a simple rescaling of the
  maximum Jeffreys' prior penalized likelihood estimator that delivers
  high accuracy in signal recovery, in terms of aggregate bias, in the
  presence of an intercept parameter. The rescaled estimator is
  effective even in cases where estimates from maximum likelihood and
  other recently proposed corrective methods based on approximate
  message passing do not exist.
\end{abstract}

\section{Introduction}

\subsection{Logistic regression with $p/n \to \kappa \in (0, 1)$}
\label{sec:pon}

Suppose we observe binary responses $y_1, \ldots, y_n$ with
$y_i \in \{0, 1\}$, and $n$ vectors of covariates $x_1, \ldots, x_n$
with $x_i \in \Re^p$. Consider a logistic regression model where
$y_1, \ldots, y_n$ are assumed to be realizations of independent
Bernoulli random variables $Y_1, \ldots, Y_n$, respectively, where
$Y_i \mid x_i$ has expectation $\mu_i$ with
\begin{equation}
  \label{eq:logistic}
  \log \frac{\mu_i}{1 - \mu_i} = \eta_i = \beta_0 + x_i^\top \beta \, ,
\end{equation}
with $\beta_0 \in \Re$ and $\beta \in \Re^p$.

Under the assumptions that $p / n \to \kappa \in (0,1)$,
$x_1, \ldots, x_n$ are generated independently from a Normal
distribution with mean $0_p$ and unknown, non-singular
variance-covariance matrix $\Sigma$, and the signal strength is
$\var(x_i^\top\beta) \to \gamma_0^2 \in (0, \infty)$,
\citet{candes+sur:2020} prove the existence of a phase transition
curve on the plane of $\kappa$ and
$\gamma = \sqrt{\beta_0^2 + \gamma_0^2}$ that characterizes when the
maximum likelihood (ML) estimate exists with probability approaching
one as $n, p \to \infty$. That phase transition is found to be sharp
for finite values of $n$ and $p$. By existence we mean that all the
components of the ML estimate are finite, and we say that the ML
estimate does not exist if at least one of its components is infinite.

\citet[Theorem~2.1]{candes+sur:2020} shows that the phase transition
curve can be computed efficiently for any given $\beta_0$ by
minimizing a function of two unknowns over a grid of values for
$\gamma$. This is in stark contrast to the usual route of detecting
infinite estimates in logistic regression, which requires the solution
of appropriate linear programs which operate only on a given set of
responses and covariates, ignoring any knowledge of the distribution
of the covariates. Such linear programs appear for example, in the
seminal work of \citet{albert+anderson:1984}, where absence of data
separation is a necessary and sufficient condition for the existence
of the ML estimate. \citet{konis:2007} provides an alternative linear
program, which is implemented in the \texttt{detectseparation} R
package \citep{detectseparation}. Solving those linear programs,
though, can be computationally demanding for large $n$ or $p$.

The phase transition curve is particularly useful because it
identifies the ($\kappa$, $\gamma$) combinations for which the
corrective methods of \citet{sur+candes:2019} and
\citet{zhao+sur+candes:2022} apply for recovering aggregate
consistency of the ML estimator and correcting inference from
likelihood-based pivots in the high-dimensional logistic regression
with $p / n \to \kappa \in (0, 1)$. Specifically for estimation,
approximate message passing arguments are used to derive a system of
nonlinear equations in \citet[expression~(5)]{sur+candes:2019}, whose
solution provides the factor by which the ML estimator needs to be
rescaled in order to recover asymptotic aggregate unbiasedness. As
\citet{sur+candes:2019} clearly note, that system of equations does
not have a unique solution if $(\kappa, \gamma)$ are on the side of
the phase transition curve where the ML estimate does not exist
asymptotically, and the corrective approaches they introduce do not
apply.

\subsection{Logistic regression with fixed $p$ and $n \to \infty$}

\citet{firth:1993} shows that for logistic regression with fixed $p$
and $n \to \infty$, the estimator of
$\theta = (\beta_0, \beta^\top)^\top$ that results by the maximization
of the Jeffreys' prior penalized log-likelihood
\begin{equation}
  \label{eq:mJPL}
  \tilde{\ell}(\theta) = \ell(\theta) + \frac{1}{2} \log \left| \bar{X}^\top W(\theta) \bar{X} \right| 
\end{equation}
has bias of order $O(n^{-2})$, which is asymptotically smaller than
the $O(n^{-1})$ bias that is expected by the ML estimator from
standard asymptotic theory \citep[see, for
example,][Section~7.3]{mccullagh:1987}.  We say ``expected'' here
because, formally, the bias of the ML estimator conditional on the
observed covariates is not defined for a logistic regression model, as
there is always at least one configuration of responses which results
in ML estimates with infinite components. In~(\ref{eq:mJPL}),
$\ell(\theta) = \sum_{i = 1}^n \{ y_i \eta_i - \log(1 +
\exp{\eta_i})\}$ is the log-likelihood of the logistic regression
model with linear predictor $\eta_i$ as in~(\ref{eq:logistic}),
$\bar{X} = \begin{bmatrix} 1_n & X \end{bmatrix}$, where $1_n$ is an
vector of $n$ ones, and $X$ is
the $n \times p$ model matrix with $x_1, \ldots, x_n$ in its rows, and
$W(\theta)$ is a diagonal matrix with $i$th diagonal element
$\mu_i (1 - \mu_i)$. Importantly, \citet{kosmidis+firth:2021} show
that the maximum Jeffreys' prior penalized likelihood (mJPL) estimator
has always finite components, even in cases where the ML estimate does
not exist, under the sole assumption that $X$ is of full rank. Those
results underpin the increasingly widespread use of mJPL and similarly
penalized likelihood estimation in logistic regression models in many
applied fields. For example, at the date of writing \citet{firth:1993}
has about $5000$ citations in Google Scholar, across areas of
scientific enquiry, with the majority being due to the properties of
mJPL in logistic regression.

Both \citet{sur+candes:2019} and \citet{kosmidis+firth:2021} also
present empirical evidence that mJPL appears to deliver a substantial
reduction in the persistent bias of the ML estimator in
high-dimensional logistic regression setting of Section~\ref{sec:pon},
whenever the ML estimate exists. We should note here that, as is the
case for ML estimation, the mJPL estimates can be computed efficiently
through a range of methods, the most popular ones being iterative
reweighted least squares \citep[see][for
details]{kosmidis+kennepagui+sartori:2020} or repeated maximum
likelihood fits on adaptively adjusted responses
\citep[see][Section~4]{kosmidis+firth:2021}. R packages such as
\texttt{brglm2} \citep{brglm2} and \texttt{logistf} \citep{logistf}
provide efficient implementations.

\subsection{Contribution}

We design and carry out a large-scale computer experiment to examine
in depth and expand the observations of \citet{sur+candes:2019} and
\citet{kosmidis+firth:2021}.

The results we gather provide further evidence that mJPL performs well
and markedly better than the ML estimator in the region below
\citet{candes+sur:2020}'s phase transition curve where the ML estimate
asymptotically exists, and tends to over-correct in the region where
the ML estimate has infinite components. A careful statistical
analysis of the results allows us to confidently state that the amount
of over-correction can be accurately characterized in terms of
$\kappa$, $\gamma$ and the relative size of the intercept parameter
$\beta_0$ and $\gamma_0$, for a wide range of $(\kappa, \gamma)$
values. As a result, we are able to formulate and contribute a
fruitful conjecture about adjusting the mJPL estimates post-fit to
almost recover, in terms of aggregate bias, the true signal to high
accuracy and for a wide range of settings. We should emphasize that
the conjecture also applies in the region above
\citet{candes+sur:2020}'s phase transition curve, where the ML
estimate does not exist with probability approaching one, and the
methods of \citet{sur+candes:2019} and \citet{zhao+sur+candes:2022} do
not apply.

\begin{conjecture}
  \label{thm:conjecture}
  Consider the logistic regression model~(\ref{eq:logistic}) where
  $x_1, \ldots, x_n$ are realizations of $n$ independent
  $N_p(0, \Sigma)$ random vectors, and assume that
  $p / n \to \kappa \in (0, 1)$ and $\beta$ is such that
  $\var(x_i^\top \beta) \to \gamma_0^2$, as $n, p \to \infty$. Then,
  the mJPL estimator $(\tilde\beta_0, \tilde\beta^\top)^\top$ from the
  maximization of~(\ref{eq:mJPL}) satisfies
  \begin{equation}
    \label{eq:aggregate_bias}
    \frac{1}{p} \sum_{j = 1}^p (\tilde\beta_j - \alpha_\star \beta_j) \overset{p}{\to} 0 \, ,
  \end{equation}
  with $\alpha_\star \le 1$. For small to moderate values of
  $\beta_0^2 / \gamma_0^2$, the scaling $\alpha_\star$ can be
  approximated by
  \[
    q(\kappa, \gamma, \gamma_0; b) = \left\{
      \begin{array}{ll}
        1  \,, & \text{if } \kappa \le h_{\rm MLE}(\beta_0, \gamma_0) \\
        \kappa^{b_1} \gamma^{b_2} \gamma_0^{b_3}\,, & \text{if } \kappa > h_{\rm MLE}(\beta_0, \gamma_0)
      \end{array}
    \right. \, ,
  \]
  where $b_1 < -1$, $b_2 < -1$, $b_3 > 0$. The function
  $h_{\rm MLE}(\beta_0, \gamma_0)$ is as defined in
  \citet[Theorem~2.1]{candes+sur:2020}, which states that the ML
  estimate exists with probability approaching one if
  $\kappa < h_{\rm MLE}(\beta_0, \gamma_0)$, and zero if
  $\kappa > h_{\rm MLE}(\beta_0, \gamma_0)$.
\end{conjecture}

Statement~(\ref{eq:aggregate_bias}) is what we refer to as aggregate
bias, and the parameters $\alpha_\star$ and
$q(\kappa, \gamma, \gamma_0; b)$ are the aggregate bias parameter, and
its approximation, respectively. Equivalently, the conjecture implies
that rescaling $\tilde{\beta}_j$ to
$\tilde{\beta}_j/ q(\kappa, \gamma, \gamma_0; b)$ produces an
estimator that has almost zero aggregate bias for small to moderate
values of $\beta_0^2 / \gamma_0^2$.

Our predictions for $b_1$, $b_2$, and $b_3$ are $-1.172$, $-1.869$,
and $0.817$, respectively; see Table~\ref{tab:conjecture}, which also
provides $95\%$ bootstrap BC$\alpha$ confidence intervals for $b_1$,
$b_2$, and $b_3$. We also have evidence that
$\kappa^{b_1} \gamma^{b_2} \gamma_0^{b_3} < 1$ when the ML estimate
does not exist asymptotically (see, for example, top left of
Figure~\ref{fig:conjecture}).

To our knowledge, there has been no other tuning-parameter free
proposal for signal recovery, in terms of aggregate bias, in
the high-dimensional logistic regression setting of
\citet{sur+candes:2019} and \citet{zhao+sur+candes:2022} with
intercept parameter, which applies regardless of whether the ML
estimate exists or not and does not require approximate message
passing machinery to operate. A proof of that conjecture, at least
when there is no intercept parameter, may proceed by adapting the
framework and results in~\citet{salehi+et+al:2019} to the case where
the likelihood is penalized by Jeffreys' prior and examining whether
the solution of the resulting system of nonlinear equations
in~\citet[see, for example, expression~(6)]{salehi+et+al:2019} is
approximately equal to the scaling factor
$q(\kappa, \gamma, \gamma_0; b)$ for small to moderate values of
$\beta_0^2 / \gamma_0^2$.

\begin{figure}[t]
  \begin{center}
    \includegraphics[width = 0.8\textwidth]{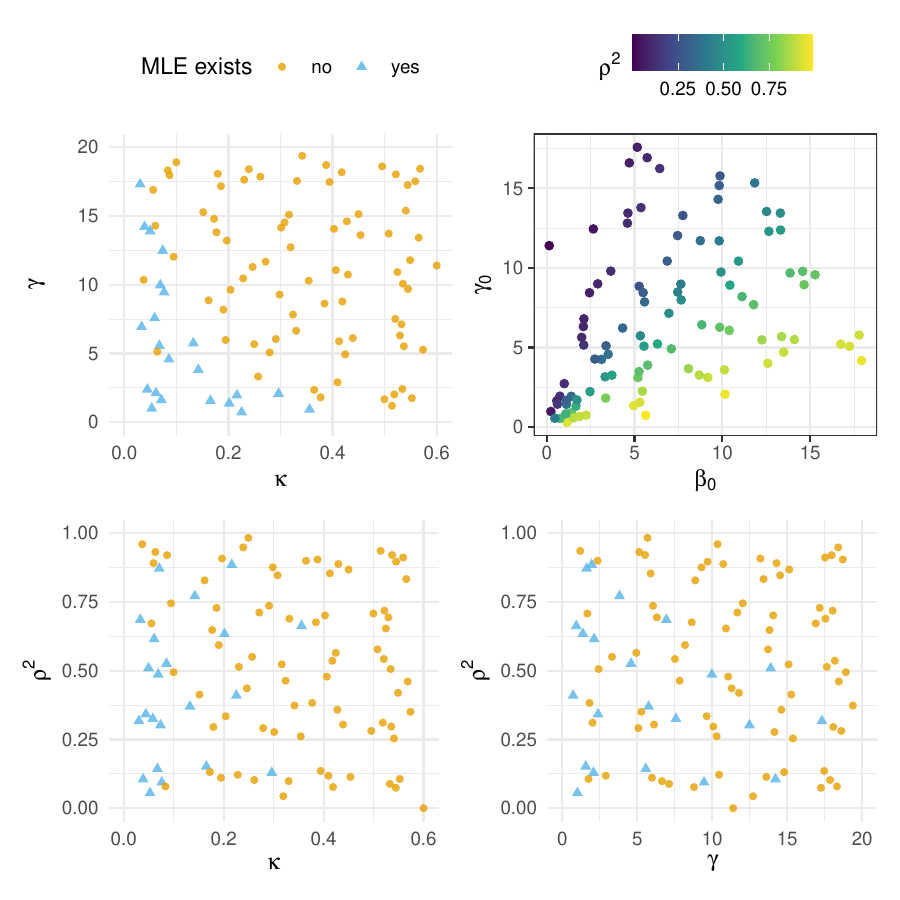}
  \end{center}
  \caption{Top left, bottom left, and bottom right: $100$ points
    $(\kappa, \gamma, \rho^2) \in (0, 0.6) \times (0, 20) \times (0,
    1)$ generated using a minimax projection design
    \citep{mak+joseph:2018}, plotted on the three two-dimensional
    subspaces. At each point, \citet[Theorem~2.1]{candes+sur:2020} is
    used to determine whether the ML estimate exists asymptotically or
    not. Top right: The implied $(\beta_0, \gamma_0)$
    combinations.}
  \label{fig:design}
\end{figure}

The computer experiment also allows us to assess the performance of
the implementation of mJPL provided by the well-used R package
\texttt{brglm2}. \texttt{brglm2} is more general than mJPL for
logistic regression, providing also methods for mean and median bias
reduction for all generalized linear models under a unifying interface
and code base \citep[see][for
details]{kosmidis+kennepagui+sartori:2020}, written exclusively in
R. Hence, it provides a good benchmark relative to a purpose-built
compiled program for mJPL for logistic regression that would naturally
be much more efficient. It is found that, in notable contrast to the
observations in \citet[Supporting Information document,
Section~D]{sur+candes:2019}, where mJPL is reported to be
computationally infeasible in high-dimensional logistic regression
settings, requiring about $2.5$ hours for $n = 2000$ and $p = 400$,
the average runtime we observed for mJPL for $n = 2000$ and $p$
ranging from $20$ to $1100$ is from milliseconds to under two
minutes. See Section~\ref{sec:performance} for details.

\section{Computer experiment}
\label{sec:highdim}

\subsection{Data generating process}
\label{sec:data_generation}
For generating a single data set from model~(\ref{eq:logistic}), under
the assumptions of Conjecture~(\ref{thm:conjecture}), we specify $n$,
$\kappa \in (0, 1)$, $\gamma_0 > 0$, $\gamma > 0$, an initial
parameter vector $\beta^*$, and set $p = \lceil n \kappa \rceil$. In
order to control the relative size of the intercept parameter and
$\gamma_0$, we define $\beta_0 = \gamma \rho$, and
$\gamma_0 = \gamma \sqrt{1 - \rho^2}$. In this way,
$\beta_0^2 / \gamma_0^2 = \rho^2 / (1 - \rho^2)$. This is the same
relationship between $\beta_0$ and $\gamma_0$ that is used for
producing \citet[Figure~1(b)]{candes+sur:2020}.

Conjecture~\ref{thm:conjecture} allows for correlated covariates as in
\citet{zhao+sur+candes:2022}. Hence, we also specify a $p \times p$,
positive definite variance-covariance matrix $\Sigma$ for the
covariates.  We should mention here that, as noted in
\citet{zhao+sur+candes:2022}, the asymptotic framework of
\citet{candes+sur:2020} is fully characterized by the values of the
dimensionality constant $\kappa$, the limiting signal strength
$\gamma_0$, and the size of the intercept parameter. The
variance-covariance matrix $\Sigma$ of the covariate vectors, while
necessary when simulating covariate vectors, can be arbitrary as long
as $\var(x_i^\top \beta) \to \gamma_0^2$. We form an $n \times p$
matrix $X$ by generating its rows from $n$ independent and identically
distributed normal random vectors with mean $0_p$ and
variance-covariance matrix $\Sigma$ having $(i,j)$th element
$\psi^{|i - j|}$, with $\psi \in (-1, 1)$. We, then, rescale
$\beta^*$ to
$\beta = \gamma_0 \beta^* / \lVert L^\top \beta^* \rVert_2$, so that
we control
$\var(x_i^\top\beta) = \beta^\top LL^\top \beta = \gamma_0^2$, where
$L$ is the Cholesky factor of $\Sigma$. Finally, we generate
$y_1, \ldots, y_n$ independently with
$Y_i \sim \text{Bernoulli}(\mu_i)$, where $\mu_i$ is specified
in~(\ref{eq:logistic}).

\subsection{Training and test phases}

The computer experiment has two phases. The first phase is a training
phase and involves a specific simulation experiment using a
space-filling design of $(\kappa, \gamma, \rho^2)$ points, and
$\psi = 0$, that is the entries of $X$ are realizations of independent
standard normal random variables. The training phase is used to generate
strong evidence for the approximation to $\alpha_\star$ in
Conjecture~\ref{thm:conjecture} and predict the value of the constants
$b_1$, $b_2$ and $b_3$.  In the test phase, the generalizability of
the prediction for the approximate form of $\alpha_\star$ is examined
using independent simulation experiments for a preset grid of
$(\kappa, \gamma, \rho^2, \psi)$, which are different from the one
used in the training phase. We produce strong evidence on the
generality of the approximate form for $\alpha_\star$ in
Conjecture~\ref{thm:conjecture}.


\section{Training phase}
\label{sec:training}

\begin{table}
  \caption{Estimates and $95\%$ confidence intervals for the
    parameters of the linear regression of $\log \delta_1^*$ on
    $\log(\kappa)$, $\log(\gamma)$ and $\log(\gamma_0)$ for all
    design points with $\rho^2 < 0.7$. The confidence intervals are
    bootstrap $\mathop{\rm BC}_\alpha$ confidence intervals computed
    using $9999$ bootstrap samples through case resampling.}
  \label{tab:conjecture}
  \begin{center}
  \begin{tabular}{lD{.}{.}{2}D{.}{.}{2}D{.}{.}{2}}
    \toprule
    & \multicolumn{1}{c}{Estimate} & \multicolumn{1}{c}{2.5 \%} & \multicolumn{1}{c}{97.5 \%} \\
    \midrule
    $b_0$ & -0.033 & -0.137 & 0.069 \\
    $b_1$ & -1.172 & -1.211 & -1.130 \\
    $b_2$ & -1.869 & -2.004 & -1.730 \\
    $b_3$ & 0.817 & 0.689 & 0.942 \\
    $\phi$ & 0.004 \\
    \bottomrule
  \end{tabular}
\end{center}
\end{table}

We compute a minimax projection design \citep{mak+joseph:2018} of
$100$ points
$(\kappa, \gamma, \rho^2) \in (0, 0.6) \times (0, 20) \times (0, 1)$
using the \texttt{minimaxdesign} R package \citep{mak:2021}. We choose
the minimax projection design because, on one hand, it has the
appealing property of uniform coverage of the design space in
worst-case scenarios, and, on the other hand, improves on uniform
coverage in projected subspaces of the design space; see
\citet{mak+joseph:2018} for more details. Figure~\ref{fig:design}
shows the computed points on all two-dimensional subspaces.
\citet[Theorem~2.1]{candes+sur:2020} is used at each of the $100$
points to determine whether the ML estimate exists asymptotically or
not. The ML estimate exists asymptotically at $22$ points.

We set $n = 2000$ and $\psi = 0$. Then, for each point on the
minimax projection design, $\beta^*$ is set to an equi-spaced grid of
length $p = \lceil n \kappa \rceil$ between $1$ and $10$, and $100$
samples of $X$ and $y$ are drawn independently as detailed in
Section~\ref{sec:data_generation}. The ML estimates of $\beta_0$ and
$\beta$ are computed only in settings where the ML estimate
asymptotically exists, and the mJPL estimates are computed everywhere
using the $\texttt{brglm\_fit()}$ method for the $\texttt{glm()}$ R
function, as provided by the $\texttt{brglm2}$ R package. We compute
the estimates so that they are accurate to the third decimal point and
allowing up to $300$ of the quasi Fisher scoring iterations
$\texttt{brglm\_fit()}$ implements; see the vignettes of
\citet{brglm2} for details on the quasi Fisher scoring iteration. The
mJPL estimates are computed starting at zero for all parameters, and
are used as starting values for the ML estimates, whenever the latter
are computed.

\begin{figure}[t]
  \begin{center}
    \includegraphics[width = 0.8\textwidth]{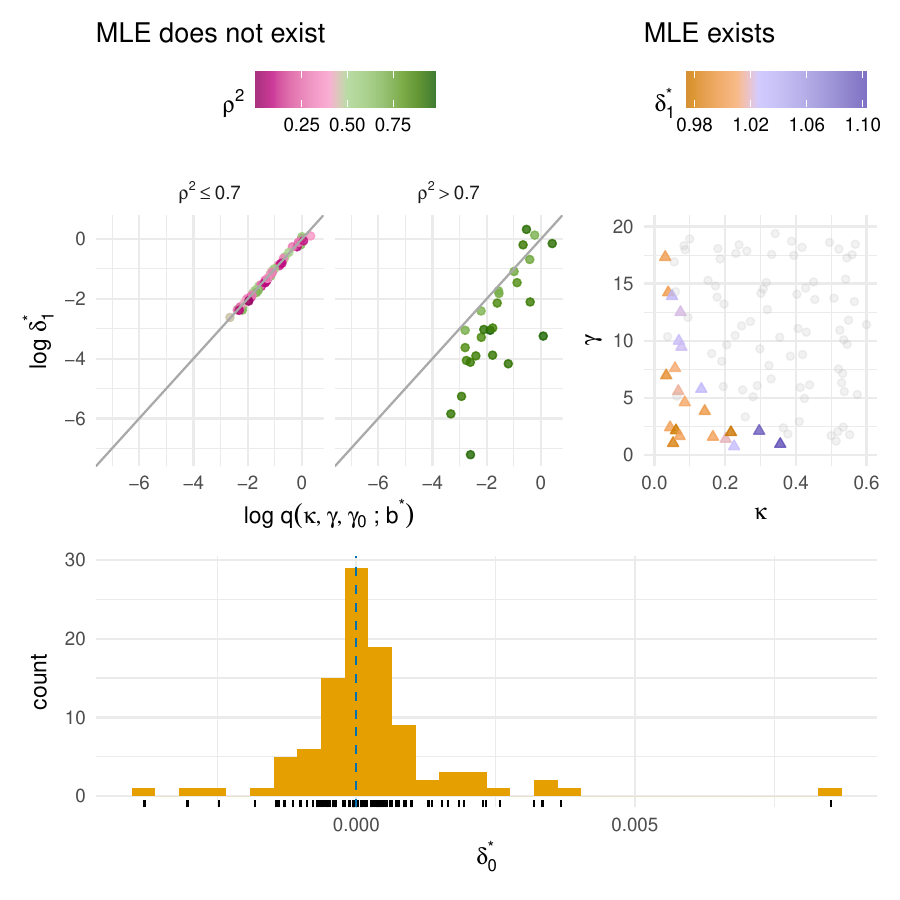}
  \end{center}
  \caption{Results from the training phase in
    Section~\ref{sec:training}. Top left: a scatterplot of
    $\log \delta_1^*$ versus
    $\log q(\kappa, \gamma, \gamma_0; b^*)$ for design points with
    $\rho^2 \le 0.7$ and design points with $\rho^2 > 0.7$, when the
    ML estimate does not exist asymptotically. The grey line is a
    $45^o$ reference line. Top right: scatterplot of the design points
    in Figure~\ref{fig:design} for which the ML estimate exists. The
    points are coloured accourding to the value of $\log
    \delta_1^*$. Bottom: Histogram of $\log \delta_1^*$ across design
    points.}
  \label{fig:conjecture}
\end{figure}

Evidence for Conjecture~\ref{thm:conjecture} is that, for all design
points $(\kappa, \gamma, \rho^2)$ (or equivalently
$(\kappa, \gamma, \gamma_0)$), the simple linear regression of the
realizations of
$\tilde\beta = (\tilde\beta_1, \ldots, \tilde\beta_p)^\top$ on the
true values $\beta = (\beta_1, \ldots, \beta_p)^\top$ has intercept
$\delta_0(\kappa, \gamma, \gamma_0) = 0$, and slope
$\delta_1(\kappa, \gamma, \gamma_0) = 1$
if the ML estimate exists asymptotically, and
$\delta_1(\kappa, \gamma, \gamma_0) = \kappa^{b_1} \gamma^{b_2}
\gamma_0^{b_3}$ if the ML estimate does not exist asymptotically.

For the case where the ML estimate does not exist, consider a
multiplicative regression model with constant coefficient of variation
\citep[see][Chapter~8]{mccullagh+nelder:1989} where
$\expect(\delta_1^*) = e^{b_o} \kappa^{b_1} \gamma^{b_2}
\gamma_0^{b_3}$. Estimates for $b_0$, $b_1$, $b_2$ and $b_3$ can be
computed using maximum likelihood for a Gamma-response generalized
linear model with log link, which is formally equivalent to
quasi-likelihood estimation \citep{wedderburn:1974} with response
variance $\phi \{\expect(\delta_1^*)\}^2$, where $\phi$ is a
dispersion parameter. The dispersion parameter can be estimated after
computing the estimates of $b_0$, $b_1$, $b_2$ and $b_3$, using a
moment-based estimator (see, for example,
\citealt[Section~8.3]{mccullagh+nelder:1989} and the
\texttt{summary.glm} function in R). In support of
Conjecture~\ref{thm:conjecture}, it is straightforward to find
moderate-valued thresholds $r \in (0, 1)$ for $\rho^2$ (or
equivalently for $\beta_0^2 / \gamma_0^2$), for which a Gamma-response
generalized linear model with log link is almost a perfect fit for all
design points with $\rho^2 \le r$. For example,
Table~\ref{tab:conjecture} shows estimates and 95\% bootstrap
confidence intervals for $b_0$, $b_1$, $b_2$ and $b_3$ based on the
$70$ design points with $\rho^2 \le r = 0.7$
($\beta_0^2 / \gamma_0^2 \le 7/3$). That fit is found to explain
$99.34 \%$ of the null deviance, and an inspection of the residuals
reveals no evidence of departures from the assumptions of a
multiplicative regression model with constant coefficient of
variation. Under the Gamma regression model,
$\expect(\log \delta_1^*) \simeq \log \mu - \phi / 2$. Given the small
value of the estimate of $\phi$, a linear regression of
$\log \delta_1^*$ on $\log(\kappa)$, $\log(\gamma)$ and
$\log(1 - \rho^2)$ should be an excellent fit, too, with almost
identical estimates to those in Table~\ref{tab:conjecture}. This is
indeed the case; that linear regression returns estimates $-0.037$,
$-1.171$, $-1.871$, and $0.822$ for $b_0$, $b_1$, $b_2$, and $b_3$,
respectively, explaining $R^2 = 99.32\%$ of the variability in
$\log\delta_1^*$.

The top left of Figure~\ref{fig:conjecture} shows a scatterplot of
$\log \delta_1^*$ versus $\log q(\kappa, \gamma, \gamma_0; b^*)$
for design points with $\rho^2 \le 0.7$ and design points with
$\rho^2 > 0.7$, where $b^* = (b_1^*, b_2^*, b_3^*)^\top$ is the
estimate of $(b_1, b_2, b_3)^\top$ given in Table~\ref{tab:conjecture}
(i.e.~ignoring $b_0$, for which there is also no evidence that it is
not zero). The points agree closely with the reference line of slope
one through the origin when $\rho^2 \le 0.7$. That agreement starts
progressively breaking down as $\rho^2$ grows, or, equivalently, as
the relative size of the intercept $\beta_0$ to the signal strength
$\gamma_0$ grows. Furthermore,
$\kappa^{b_1^*} \gamma^{b_2^*} \gamma_0^{b_3^*} < 1$, except from a few
points that are close to phase transition.

At the top right of Figure~\ref{fig:conjecture} we see that the values
of $\delta_1^*$ at the design points where the ML estimate exists
asymptotically are about one (having a sample mean of $1.016$ and
standard deviation of $0.0311$). This is strong evidence for the
conjectured approximation to the aggregate bias parameter
$\alpha_\star$ when the ML estimate exists. The finite sample
distribution of $\delta_1^*$ for $n = 2000$, when the ML estimate
asymptotically exists, seems to be right-skewed (sample skewness is
$1.483$), with larger values at points close to the phase transition
for existence.

Finally, as is also apparent from the histogram on the bottom of
Figure~\ref{fig:conjecture}, the distribution of $\delta_0^*$ is
concentrated around zero with minuscule variance.

\section{Test phase}
\label{sec:test}

\begin{figure}[hbtp]
  \begin{center}
    \includegraphics[width = 0.8\textwidth]{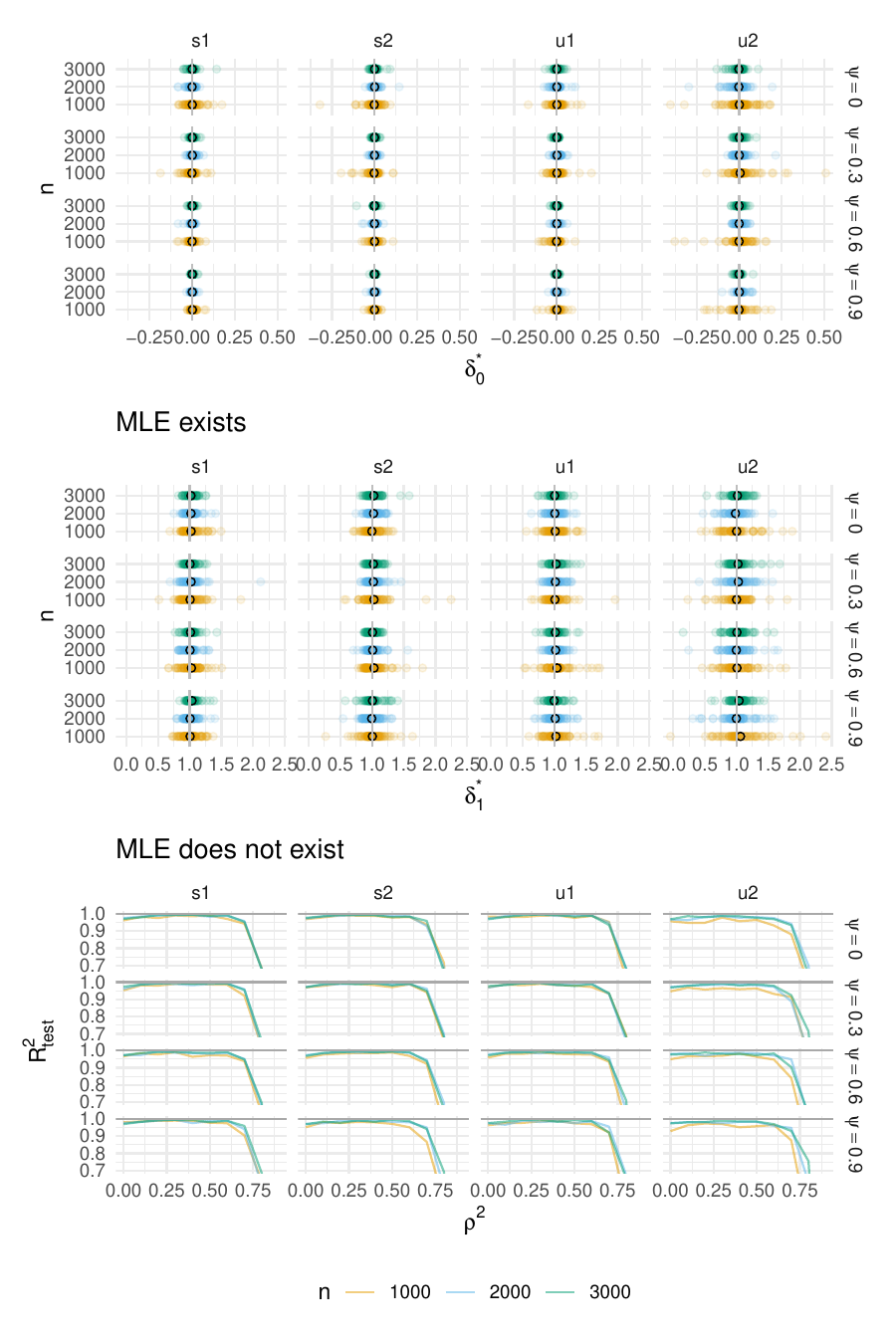}
  \end{center}
  \caption{Results from the test phase in Section~\ref{sec:test}.  Top
    and middle: Strip plots of the estimates $\delta_{0, {\rm test}}$,
    and $\delta_{1, {\rm test}}$ when the ML estimate exists
    asymptotically, respectively, for all combinations of $n$, $\psi$,
    and $\beta^*$ configuration. The black circles are the empirical
    averages of the values displayed in each strip plot. Bottom:
    $R^2_{\rm test}$, when the ML estimate does not exist
    asymptotically, as a function of $\rho^2$ for all combinations of
    $n$, $\psi$, and $\beta^*$ configuration.}
  \label{fig:test}
\end{figure}

\begin{figure}[hbtp]
  \begin{center}
    \includegraphics[width = 0.8\textwidth]{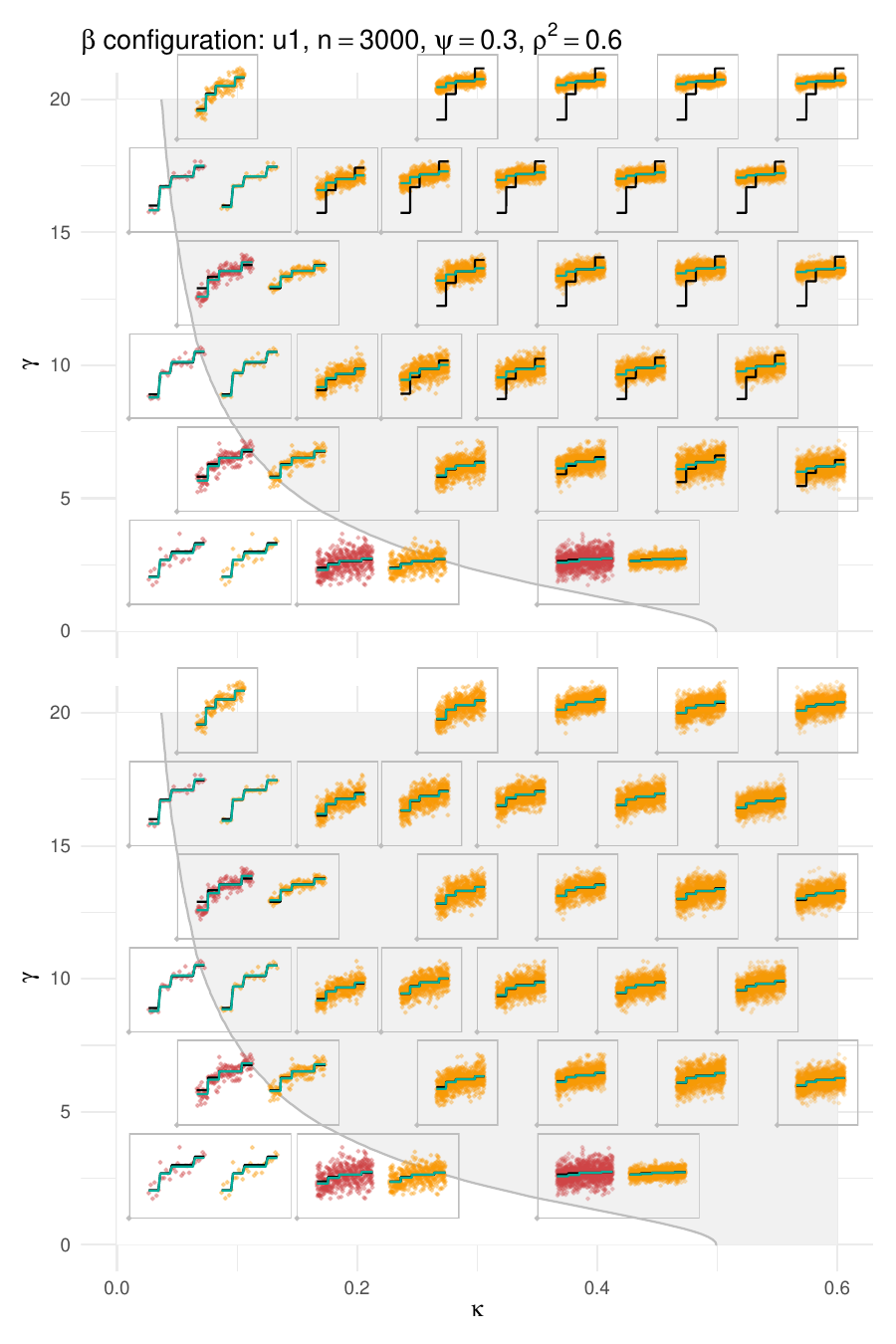}
  \end{center}
  \caption{mJPL estimates (top) and their rescaled versions after
    division by~(\ref{eq:prediction}) (bottom) for $n = 3000$,
    $\psi = 0.3$, $\rho^2 = 0.6$, and $\beta^*$ configuration u1
    versus the parameter index, with one sample per $(\kappa, \gamma)$
    point. The ML estimate exists asymptotically in the white area and
    does not exist asymptotically in the grey area. The blue segments
    show the sample mean of the estimates for each value of the truth,
    and the black segments are the truth.}
  \label{fig:illustration1}
\end{figure}

We now test our prediction of the approximation to the aggregate bias parameter
\begin{equation}
  \label{eq:prediction}
  q(\kappa, \gamma, \gamma_0; b^*) =
  \left\{\begin{array}{ll}
           1 \,, & \text{if } \kappa \le h_{\rm MLE}(\beta_0, \gamma_0) \\
           \kappa^{b_1^*} \gamma^{b_2^*} \gamma_0^{b_3^*}\,, & \text{if } \kappa > h_{\rm MLE}(\beta_0, \gamma_0)
 \end{array}\right.\,,
\end{equation}
developed in Section~\ref{sec:training}, using independent simulation
experiments. We consider all possible combinations of
$n \in \{1000, 2000, 3000\}$, $\psi \in \{0, 0.3, 0.6, 0.9\}$,
$\rho^2 \in \{0, 0.1, \ldots, 0.9\}$, and the following four
configurations for $\beta^*$:
\begin{itemize}
\item[s1.] an equi-spaced grid of length $p$ between $-10$ and $10$,
\item[s2.] the first $\lceil p / 5 \rceil$ elements
  are $-10$, the second $\lceil p / 5 \rceil$ elements are $10$, and the
  remaining elements are zero,
\item[u1.] the first $\lceil p / 5 \rceil$
  elements are $-3$, the second $\lceil p / 5 \rceil$ elements are -1,
  the last $\lceil p / 5 \rceil$ elements are 1, and all other elements
  are zero,
\item[u2.] an equi-spaced grid of length $p$ between $1$ and $10$,
\end{itemize}
where $p = \lceil n \kappa \rceil$. Configurations s1 and s2 are
symmetric about zero, while u1 and u2 are not. For each of
the $480$ combinations, we simulate one sample, as detailed in
Section~\ref{sec:data_generation}, at each of the $30$
$(\kappa, \gamma)$ points shown in the panels of
Figure~\ref{fig:illustration1}. We denote the set of those points as
$D_{\rm test}$. Then, from each sample, we estimate the intercept
$\delta_0$ and slope $\delta_1$ of the simple linear regression of the
mJPL estimates on the corresponding true values.

The top and middle panels of Figure~\ref{fig:test} show strip plots of
the estimates $\delta_{0, {\rm test}}$ for all settings, and the
estimates $\delta_{1, {\rm test}}$ when the ML estimate exists
asymptotically, respectively, for all combinations of $n$, $\psi$, and
$\beta^*$ configuration. As is evident, those estimates are
concentrated around zero and one, respectively, in all test cases,
with variance decreasing with $n$, exactly as we would expect under
Conjecture~\ref{thm:conjecture}. The quality of the predictions
$q(\kappa, \gamma, \gamma_0; b^*) =
\kappa^{b_1^*}\gamma^{b_2^*}\gamma_0^{b_3^*}$, with $b^*$ as in
Table~\ref{tab:conjecture}, for $\delta_{1, {\rm test}}$ when the ML
estimate does not exist asymptotically can be assessed for each
$\beta^*$ configuration, using the out-of-sample version of the
coefficient of determination
\[
  R^2_{{\rm test}}(n, \rho^2, \psi) = \frac{\sum \left\{\log \delta_{1, {\rm test}} - \log
      q(\kappa, \gamma, \gamma_0; b^*)\right\}^2}{\sum \left\{\log \delta_{1, {\rm test}} -
      \sum \log \delta_{1, {\rm
          test}} / 30 \right\}^2} \, .
\]
All summations are over all $(\kappa, \gamma) \in D_{\rm test}$, and
the dependence of $\delta_{1, {\rm test}}$ on $n$, $\psi$, $\kappa$,
$\gamma$, and $\rho^2$ (or equivalently $\gamma_0$) is suppressed for
notational convenience. The quantity
$R^2_{{\rm test}}(n, \rho^2, \psi) \in (-\infty, 1]$, and a value of
$1$ has the usual interpretation that the predicted values
$\log q(\kappa, \gamma, \gamma_0; b^*)$ explain all the variability in
$\log \delta_{1, {\rm test}}$. The bottom panel of
Figure~\ref{fig:test} shows the value of $R^2_{\rm test}$ as a
function of $\rho^2$ for all combinations of $n$ and $\psi$ we
consider. The approximation $q(\kappa, \gamma, \gamma_0; b^*)$ to the
aggregate bias parameter $\alpha_\star$ is found to have excellent
out-of-sample performance for small to moderate values of $\rho^2$, as
we would expect if Conjecture~\ref{thm:conjecture} holds.

\begin{figure}[hbtp]
  \begin{center}
    \includegraphics[width = 0.8\textwidth]{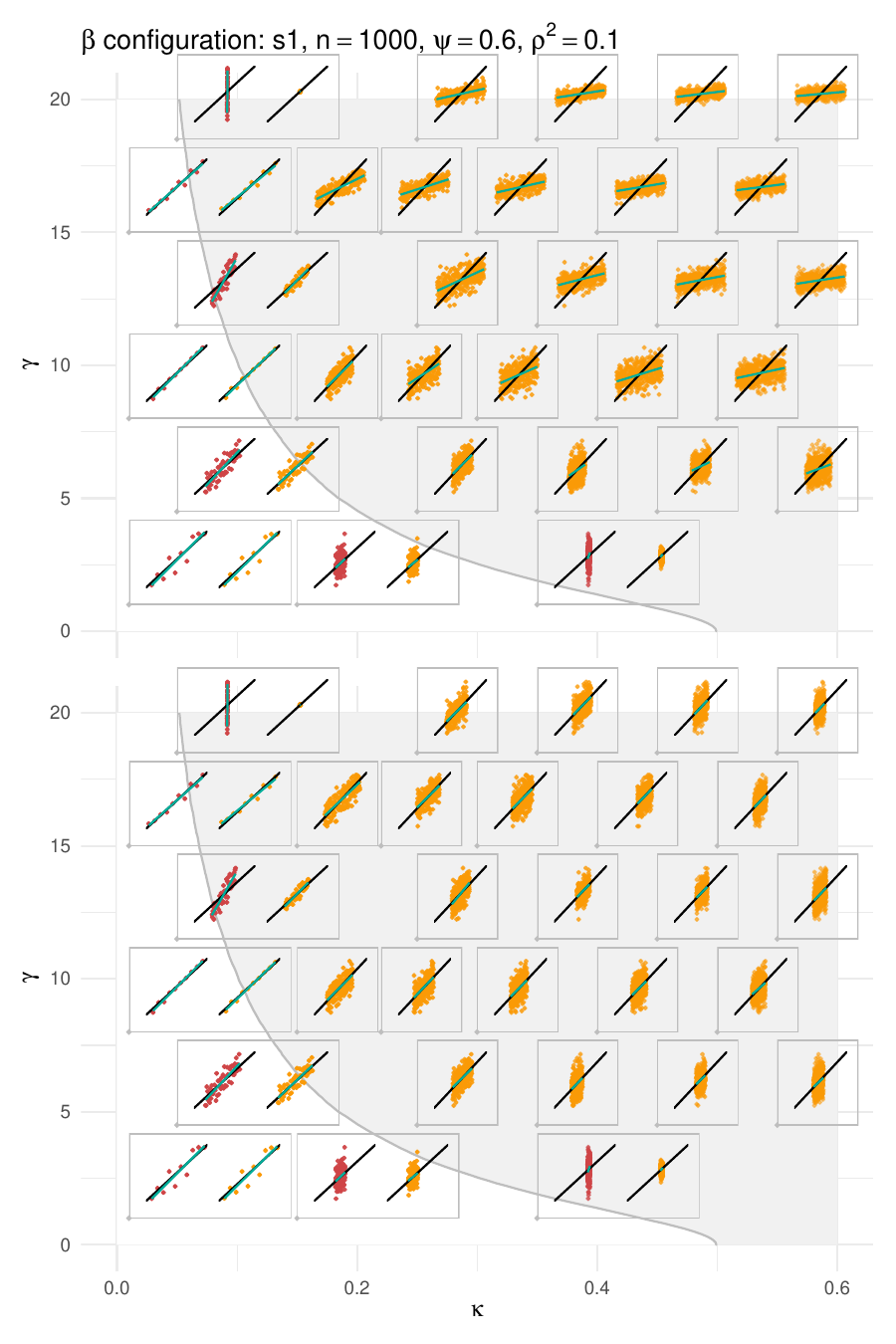}
  \end{center}
  \caption{mJPL estimates (top) and their rescaled versions after
    division by~(\ref{eq:prediction}) (bottom) for $n = 1000$,
    $\psi = 0.6$, $\rho^2 = 0.1$, and $\beta^*$ configuration s1
    versus $\beta$, with one sample per $(\kappa, \gamma)$ point. The
    ML estimate exists asymptotically in the white area and does not
    exist asymptotically in the grey area. The blue lines are from a
    simple linear regression of the estimates on $\beta$, and the
    black lines are $45^o$ reference lines.}
  \label{fig:illustration2}
\end{figure}

Figure~\ref{fig:illustration1} shows the mJPL estimates and their
rescaled versions after division by
$q(\kappa, \gamma, \gamma_0; b^*)$, for $n = 3000$, $\psi = 0.3$,
$\rho^2 = 0.6$, and $\beta^*$ configuration u1, and
Figure~\ref{fig:illustration2}, for $n = 1000$, $\psi = 0.6$,
$\rho^2 = 0.1$, and $\beta^*$ configuration s1, with one sample per
$(\kappa, \gamma)$ point. The rescaled mJPL estimator results in
excellent signal recovery.

\section{Computational performance}
\label{sec:performance}

\begin{table}[h!] 
  \caption{Summaries on computational performance from using the
    \texttt{brglm2} R package for mJPL on data sets from the test
    phase for $n = 2000$ and $\beta^*$ configuration s2, on a 2021
    MacBook Pro with Apple M1 Max chip and 64 GB RAM. Summaries are
    over all combinations of $\psi \in \{0, 0.3, 0.6, 0.9\}$ and
    $\rho^2 \in \{0, 0.1, \ldots, 0.9\}$, for each $(\kappa, \gamma)$
    point on the panels of Figure~\ref{fig:illustration1}. $t$ is
    average computational time in minutes, and $\min k$, $\bar{k}$,
    and $\max k$ are minimum, average and maximum required number of
    iterations, respectively).  Computational times less than $0.01$
    minutes are shown as $0.00$.}
  \begin{center}
  \begin{tabular}{D{.}{.}{2}D{.}{.}{2}D{.}{.}{0}D{.}{.}{2}D{.}{.}{0}D{.}{.}{2}D{.}{.}{0}}
    \toprule
       \multicolumn{1}{c}{$\kappa$} & \multicolumn{1}{c}{$\gamma$} & \multicolumn{1}{c}{$p$} & \multicolumn{1}{c}{$t$} & \multicolumn{1}{c}{$\min k$} & \multicolumn{1}{c}{$\bar{k}$} & \multicolumn{1}{c}{$\max k$} \\
\midrule
0.01 &  1.00 &   20 & 0.00 &  3 &  3.00 &  3 \\
0.01 &  8.00 &   20 & 0.00 &  7 &  8.82 & 15 \\
0.01 & 15.00 &   20 & 0.00 &  9 & 14.00 & 50 \\
0.05 &  4.50 &  100 & 0.01 &  6 &  7.53 & 11 \\
0.05 & 11.50 &  100 & 0.02 & 11 & 19.00 & 58 \\
0.05 & 18.50 &  100 & 0.04 & 16 & 33.58 & 96 \\
0.15 &  1.00 &  300 & 0.06 &  3 &  3.10 &  4 \\
0.15 &  8.00 &  300 & 0.40 & 10 & 39.50 & 80 \\
0.15 & 15.00 &  300 & 0.33 &  8 & 29.55 & 47 \\
0.22 &  8.00 &  440 & 0.64 &  7 & 28.32 & 60 \\
0.22 & 15.00 &  440 & 0.69 &  5 & 21.85 & 41 \\
0.25 &  4.50 &  500 & 1.45 & 11 & 39.02 & 77 \\
0.25 & 11.50 &  500 & 0.75 &  5 & 20.43 & 47 \\
0.25 & 18.50 &  500 & 0.73 &  5 & 18.10 & 35 \\
0.30 &  8.00 &  600 & 1.07 &  5 & 19.20 & 42 \\
0.30 & 15.00 &  600 & 0.83 &  4 & 15.03 & 26 \\
0.35 &  1.00 &  700 & 0.51 &  5 &  5.95 &  7 \\
0.35 &  4.50 &  700 & 1.33 &  7 & 25.15 & 58 \\
0.35 & 11.50 &  700 & 0.88 &  6 & 13.95 & 24 \\
0.35 & 18.50 &  700 & 0.83 &  6 & 13.78 & 23 \\
0.40 &  8.00 &  800 & 1.39 &  8 & 13.55 & 21 \\
0.40 & 15.00 &  800 & 1.20 &  8 & 12.35 & 19 \\
0.45 &  4.50 &  900 & 1.98 &  9 & 17.62 & 38 \\
0.45 & 11.50 &  900 & 1.48 &  8 & 11.57 & 19 \\
0.45 & 18.50 &  900 & 1.41 &  7 & 11.30 & 18 \\
0.50 &  8.00 & 1000 & 1.71 &  9 & 11.80 & 21 \\
0.50 & 15.00 & 1000 & 1.68 &  9 & 11.15 & 17 \\
0.55 &  4.50 & 1100 & 1.76 & 11 & 14.00 & 25 \\
0.55 & 11.50 & 1100 & 1.74 &  9 & 11.57 & 17 \\
0.55 & 18.50 & 1100 & 1.66 &  9 & 11.70 & 15 \\
\bottomrule
  \end{tabular}
\end{center}
\label{tab:performance}
\end{table}

\citet[Supporting Information document, Section~D]{sur+candes:2019}
reports that mJPL has been computationally infeasible for high
dimensions, with a runtime of approximately 10 minutes for $n = 1000$
and $p = 200$, and of over 2.5 hours for $n = 2000$ and $p = 400$. In
notable contrast to those observations, for $n = 2000$, and
configuration s2, the average runtime for mJPL using the
\texttt{brglm2} R package ranges from milliseconds to just below 2
minutes, with no particular care in the choice of starting
values. Averages are computed using data sets from the test phase and
over all combinations of $\psi \in \{0, 0.3, 0.6, 0.9\}$,
$\rho^2 \in \{0, 0.1, \ldots, 0.9\}$ for each $(\kappa, \gamma)$ point
on the panels of
Figure~\ref{fig:illustration1}. Table~\ref{tab:performance} provides
detailed summaries on computational performance.

Purpose-build implementations for mJPL can result in dramatic reduction in
computational times. For example, the repeated ML fits on adjusted
responses and totals in \citet[Section~4]{kosmidis+firth:2021}, paired
with compiled programs for ML estimation of logistic regression
models will result in substantially better runtimes.

\section{Concluding remarks}

\subsection{Covariate distribution}
\label{sec:bernoulli}

\begin{figure}[hbtp]
  \begin{center}
    \includegraphics[width = 0.8\textwidth]{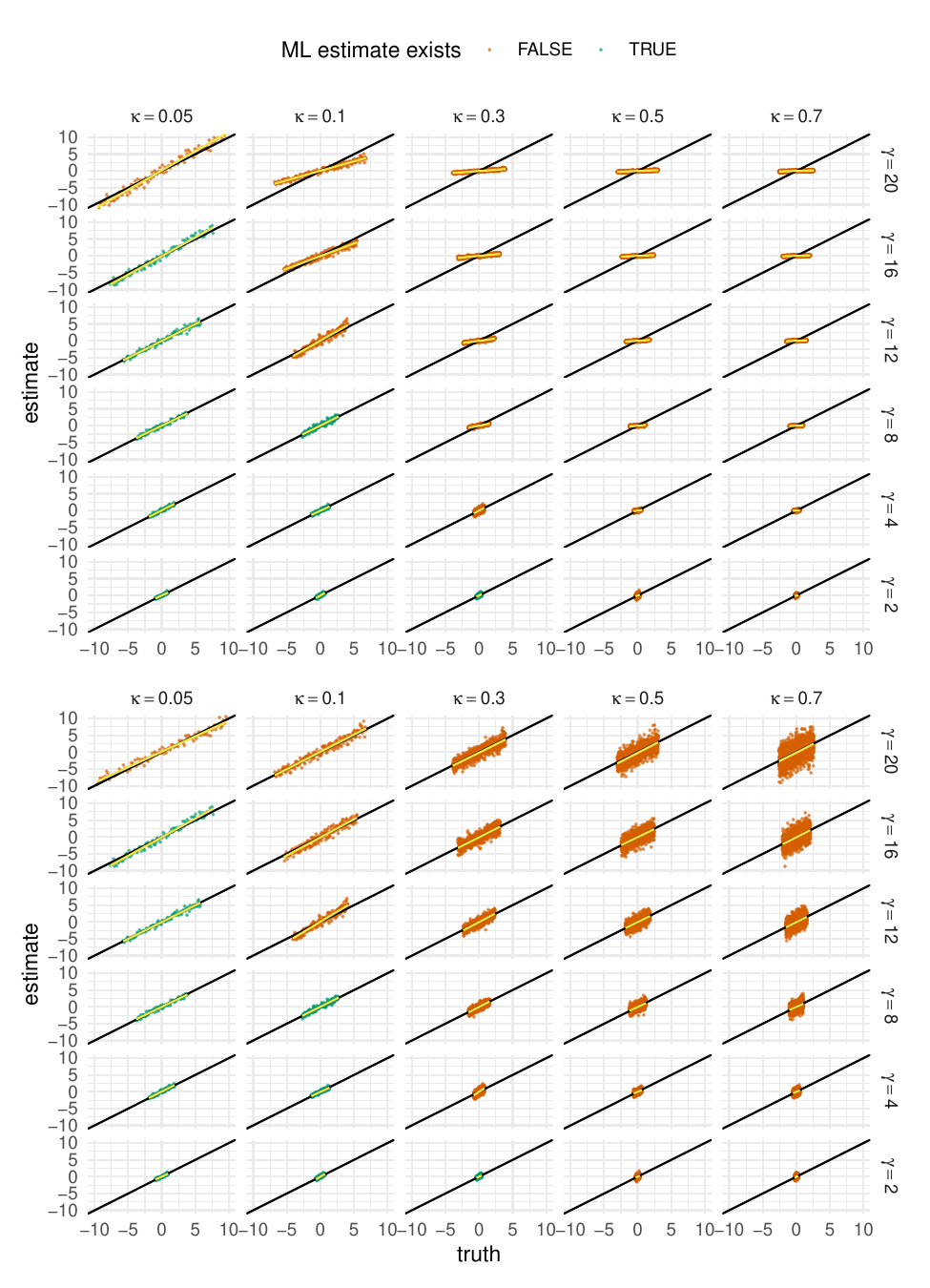}
  \end{center}
  \caption{mJPL estimates (top) and their rescaled versions after
    division by~(\ref{eq:prediction}) (bottom) for $n = 2000$,
    $\lambda = 0.1$, $\rho^2 = 0.3$, and $\beta^*$ configuration s1,
    with one sample per $(\kappa, \gamma)$ combination in the setting
    of Section~\ref{sec:bernoulli}. Existence (green) or not (orange)
    of the ML estimate is determined using the
    \texttt{detectseparation} R package for each sample. The yellow
    lines are from a simple linear regression of the estimates
    on $\beta$, and the black lines are $45^o$ reference lines.}
  \label{fig:bernoulli}
\end{figure}

The numerical evidence in \citet[Section~6]{zhao+sur+candes:2022}
illustrate that their theoretical results for multivariate normal
covariates continue to apply for a broad class of covariate
distributions with sufficiently light tails. Our empirical experience
is that the same holds for Conjecture~\ref{thm:conjecture}! For
example, consider a setting where $x_i$ consists of independent
Bernoulli random variables with probability of success
$\lambda = 0.1$, and $\beta^*$ configuration s1. For each combination
of $\kappa \in \{0.05, 0.1, 0.3, 0.5, 0.7\}$ and
$\gamma \in \{2, 4, 8, 12, 16, 20\}$, and $\rho^2 = 0.3$, we generate
a data set with $n = 2000$ as in Section~\ref{sec:data_generation},
with the only difference that
$\beta = \gamma_0 \beta^* / (\sqrt{\lambda (1 - \lambda)}\lVert L^\top
\beta^* \rVert_2)$, so that $\var(x_i^\top \beta) = \gamma_0^2$ for
independent Bernoulli covariates. In absence of asymptotic theory for
the existence of the ML estimate when the covariates are realizations of
Bernoulli random variables, we determine whether the ML estimate
exists for each simulated data set using the \texttt{detectseparation}
R package.

Figure~\ref{fig:bernoulli} shows the mJPL estimates and their rescaled
versions after division by the approximation to the aggregate bias
parameter in~(\ref{eq:prediction}) versus $\beta$, where the
conditions $\kappa < h_{\rm MLE}(\beta_0, \gamma_0)$ and
$\kappa > h_{\rm MLE}(\beta_0, \gamma_0)$ are replaced by whether the
ML estimate exists or not, respectively, at each data set. We should
emphasize that our prediction~(\ref{eq:prediction}) comes from
experiments involving model matrices that consist of realizations of
independent standard normal random variables. We again observe
excellent signal recovery. As in the case of normal covariates, we
found that the quality of the approximation
$q(\kappa, \gamma, \gamma_0; b^*)$ to $\alpha_\star$ progressively
deteriorates as $\rho^2$ increases beyond about $0.7$.

\subsection{Aggregate mean squared error}
\label{sec:ridge}

Another estimator that can result in finite estimates beyond the phase
transition curve, is the logistic ridge regression estimator, which is
defined as the maximizer of
$\ell(\beta) / n - \lambda \sum_{j=1} \beta_{j}^2 / (2p)$
$(\lambda \geq 0)$.  \citet[Theorem~3]{salehi+et+al:2019} allows to
derive a rescaled ridge estimator with zero aggregate bias when
$p / n \to \kappa \in (0, 1)$ and $\beta_0 = 0$, and characterize its
asymptotic aggregate mean squared error (aMSE), for any value of the
tuning parameter $\lambda$ \citep[see,
also,][Chapter~4]{sur:2019}. Hence, for given $\kappa$ and $\gamma$,
we can identify the value of $\lambda$ that results in the minimum
asymptotic aMSE ridge estimator with zero asymptotic aggregate bias
when $p / n \to \kappa \in (0, 1)$.

\begin{figure}[t]
  \begin{center}
    \includegraphics[width = 0.9\textwidth]{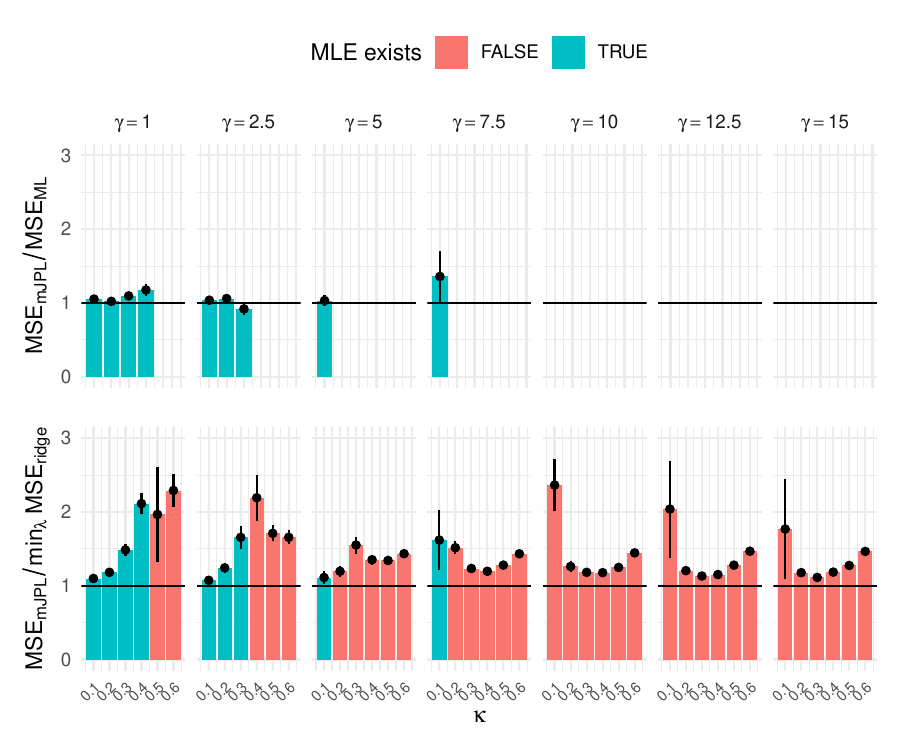}
  \end{center}
  \caption{The average estimated aMSE of the rescaled mJPL estimator
    from $50$ independent samples at each of the $42$ combinations of
    $\kappa \in \{0.1, 0.2, \ldots, 0.6\}$ and
    $\gamma \in \{1, 2.5, 5.0, \ldots, 15.0\}$ from the simulation
    experiment of Section~\ref{sec:ridge}. The average estimated aMSE
    are reported relative to the asymptotic aMSE of the rescaled ML
    estimator of \citet{sur+candes:2019}, whenever that exists (top),
    and the minimum asymptotic aMSE rescaled ridge estimator of
    \citet{salehi+et+al:2019} (bottom). The black vertical lines are
    $\pm 3$ estimated standard deviations from the estimated relative aMSE.}
  \label{fig:vs-ridge}
\end{figure}

Assuming that Conjecture~\ref{thm:conjecture} is accurate, we conduct
a simulation study to estimate the aMSE of the rescaled mJPL estimator
using the approximation to the aggregate bias parameter
in~(\ref{eq:prediction}), and compare it to the asymptotic aMSE of the
rescaled ridge estimator.

For $n = 2000$ and $p = n \kappa$, we adapt the no-intercept
($\gamma = \gamma_0$) setup of \citet{salehi+et+al:2019}, and simulate
$p$-dimensional covariate vectors $x_i$ with entries that are
realizations of independent normal random variables with mean 0 and
variance $1 / p$. We simulate response values $y_1, \ldots, y_n$
independently, where $y_i$ is a realization of a Bernoulli random
variable with probability $(1 + \exp(-x_i^\top \beta))^{-1}$, where
$\beta$ is a rescaled version of configuration s1 in
Section~\ref{sec:test} so that $\|\beta \|_2^2 / p = \gamma^2$.

For each of the $42$ combinations of
$\kappa \in \{0.1, 0.2, \ldots, 0.6\}$ and
$\gamma \in \{1, 2.5, 5.0, \ldots, 15.0\}$, we simulate $50$ datasets
and for each data set we estimate the aMSE
\[
  \frac{1}{p} \sum_{j = 1}^p\left\{\tilde{\beta}_j / q(\kappa, \gamma, \gamma; b^*) - \beta_j\right\}^2 \
\]
of the rescaled mJPL estimates. The aggregate biases of the rescaled
mJPL estimator across $(\kappa, \gamma)$ combinations are found to be
small in absolute value ranging between $-0.062$ and $0.076$, which is
also evidence for the approximation to the aggregate bias parameter in
Conjecture~\ref{thm:conjecture}. Figure~\ref{fig:vs-ridge} shows the
ratio of the average estimated aMSE of the rescaled mJPL estimator
over the asymptotic aMSE of the rescaled ML estimator of
\citet{sur+candes:2019}, and over the minimum asymptotic aMSE rescaled
ridge estimator of \citet{salehi+et+al:2019}. The rescaled mJPL
estimator appears to have overall similar performance to the ML
estimator, when the latter exists. The performance of the rescaled
mJPL estimator is found to be close to the minimum aMSE rescaled ridge
estimator for moderate values of $\kappa$ and gets increasingly closer
as $\gamma$ increases. Notably, the worst performances with relative
aMSE value around 2, correspond to cases with $(\kappa, \gamma)$ close
to the phase transition curve for the existence of the ML
estimate. This is when the conjectured aggregate bias
parameter~\ref{eq:prediction} transitions from $1$ to a function of
$\kappa$ and $\gamma$. This may either be the actual theoretical
behaviour of the mJPL estimator or evidence that a more refined
approximation to aggregate bias parameter in
Conjecture~\ref{thm:conjecture} is required close to the phase
transition.

\subsection{Estimating constants in Conjecture~\ref{thm:conjecture}}

If Conjecture~\ref{thm:conjecture} holds, even with our predictions
$b^*$ for $b$, its use in real data applications relies on determining
$\kappa$, $\gamma$, $\beta_0$, and the existence of the ML
estimate. The existence of the ML estimate can be established for any
given data set using the linear program in \citet{konis:2007}, as is
implemented in the \texttt{detectseparation} R package, and $\kappa$
can be estimated as $p / n$. For multivariate normal covariates and
potentially other distributions with light tails, when the MLE exists,
the procedure in \citet[Section~7.2]{zhao+sur+candes:2022} can be
readily used to get estimates for $\beta_0$ and $\gamma_0$. For a
procedure that operates regardless if the MLE exists or not,
adaptations of the SLOE procedure of \citet{yadlowsky+etal:2021} is
a promising direction.

\section{Supplementary material}
The repository at
\url{https://github.com/ikosmidis/mJPL-conjecture-supplementary}
provides code to reproduce the experiments, figures, and tables in the
current paper. We also provide R image files with the outputs of all
experiments.

\section{Declarations}

For the purpose of open access, the authors have applied a Creative
Commons Attribution (CC BY) licence to any Author Accepted Manuscript
version arising from this submission.

\bibliographystyle{jss2}
\bibliography{mJPL-conjecture}

\end{document}